\documentclass[aps,pre,onecolumn,showpacs,superscriptaddress,groupedaddress]{revtex4}  
\usepackage{graphicx}  
\usepackage{dcolumn}   
\usepackage{bm}        
\usepackage{amssymb}   
\usepackage{amsmath}   
\usepackage{dsfont}    

\usepackage[english]{babel}
\usepackage{pgfplots}
\usepackage{textcomp}
\usepackage{color}

\begin{document}

\widetext


\title{Formation and dynamics of a semi vortex ring connected to a free surface}

\author{Alexandre~Vilquin}
\email[To whom correspondence should be addressed. E-mail: ]{vilquin.alexandre@laposte.net}
\affiliation{Gulliver, UMR CNRS 7083, PSL Research University, ESPCI-Paris, 75005 Paris, France}
\affiliation{Laboratoire de Physique et M\'ecanique des Milieux H\'et\'erog\`enes, UMR CNRS 7636, ESPCI-Paris, PSL Research University, Sorbonne Universit\'e, Universit\'e Paris Diderot, 75231 Paris CEDEX 5, France}

\author{Samuel~Hidalgo-Caballero}
\affiliation{Gulliver, UMR CNRS 7083, PSL Research University, ESPCI-Paris, 75005 Paris, France}

\author{Vincent~Pagneux}
\affiliation{Laboratoire d'Acoustique de l'Universit\'e du Maine, UMR CNRS 6613, 72085 Le Mans CEDEX 9, France}

\author{Agn\`es~Maurel}
\affiliation{Institut Langevin, UMR CNRS 7587, ESPCI-Paris, 75005 Paris, France}

\author{Philippe~Petitjeans}%
\affiliation{Laboratoire de Physique et M\'ecanique des Milieux H\'et\'erog\`enes, UMR CNRS 7636, ESPCI-Paris, PSL Research University, Sorbonne Universit\'e, Universit\'e Paris Diderot, 75231 Paris CEDEX 5, France}


\begin{abstract}
Playing a role in the locomotion of some animals such as the water strider, the formation and dynamics of a semi vortex ring connected to a free surface are experimentally investigated. This semi vortex ring is generated by the circular motion of a flat circular disk in water. Digital Particle Image Velocimetry provides velocity fields and vortex properties. We show how in a broad range of Reynolds numbers, the properties of the semi vortex rings are related to the disk characteristics. In particular, our results highlight a formation process more complex than for a complete vortex ring produced by a piston stroke. In addition to the classical rolling up at the rear of the disk, a shedding phenomenon occurs on the leading edge, producing secondary vortices. The Strouhal number related to this shedding process reveals that it comes from the free-shear instability of the boundary layer identified in the near wake of a cylinder by Bloor. Finally we show that the disk thickness affects the final properties of the semi vortex rings through emission frequency of secondary vortices.
\end{abstract}

\maketitle

\section{Introduction}
Vortex rings are torus-shaped vortices that exist in many different situations in fluid mechanics. They are involved in the locomotion of numerous animals, like birds \cite{Kokshaysky1979} and fish \cite{Lauder2002}. By studying biological examples and experiments, Dabiri \cite{Dabiri2009} showed how the understanding of the vortex formation is essential to optimize the propulsion in engineering. Vortex rings also play a major role for blood pumping in the left ventricle \cite{Arvidsson2016} and are considered as an index of cardiac health in the human heart \cite{Gharib2006}. At very small scales, Rayfield \cite{Rayfield1964} have identified the presence of charged vortex rings with a quantized circulation in superfluid helium. A comprehensive review about vortex ring properties is given by Shariff and Leonard \cite{Shariff1992}.

Because of their prevalence, many experimental and numerical studies have been devoted to the vortex ring properties such as the formation process, the displacement velocity and the dynamics of the core circulation. Vortex rings can be produced in several ways. A first simple example is a liquid droplet falling onto a free surface. After impacting the surface, the droplet induces a vortex ring which propagates in the liquid \cite{SanLee2015}. However, the most common way to generate vortex rings in laboratory is by moving a column of liquid with a circular piston through an orifice. With a dye in liquid, Didden \cite{Didden1979} highlighted how the boundary layer separates and rolls up into a vortex at the edges of the orifice. He proposed a vorticity-flux model to calculate the vortex ring circulation. Digital Particle Image Velocimetry (DPIV) has allowed numerous quantitative studies on vortex rings generated by a piston stroke \cite{Willert1991}. With this method, Gharib \emph{et al.} \cite{Gharib1998} showed that the maximum circulation is reached for a characteristic dimensionless time during the vortex ring formation. Glezer \cite{Glezer1988} established a link between the formation process and the laminar or turbulent nature of a vortex ring. Weigand \cite{Weigand1997} studied the evolution of vorticity distribution and velocity displacement for laminar vortex rings during propagation from models proposed by Didden \cite{Didden1979}, Pullin \cite{Pullin1979} and Saffman \cite{Saffman1992}.

\begin{figure}\centering
	\includegraphics[width=12cm,trim=0 0 0 0,clip=true]{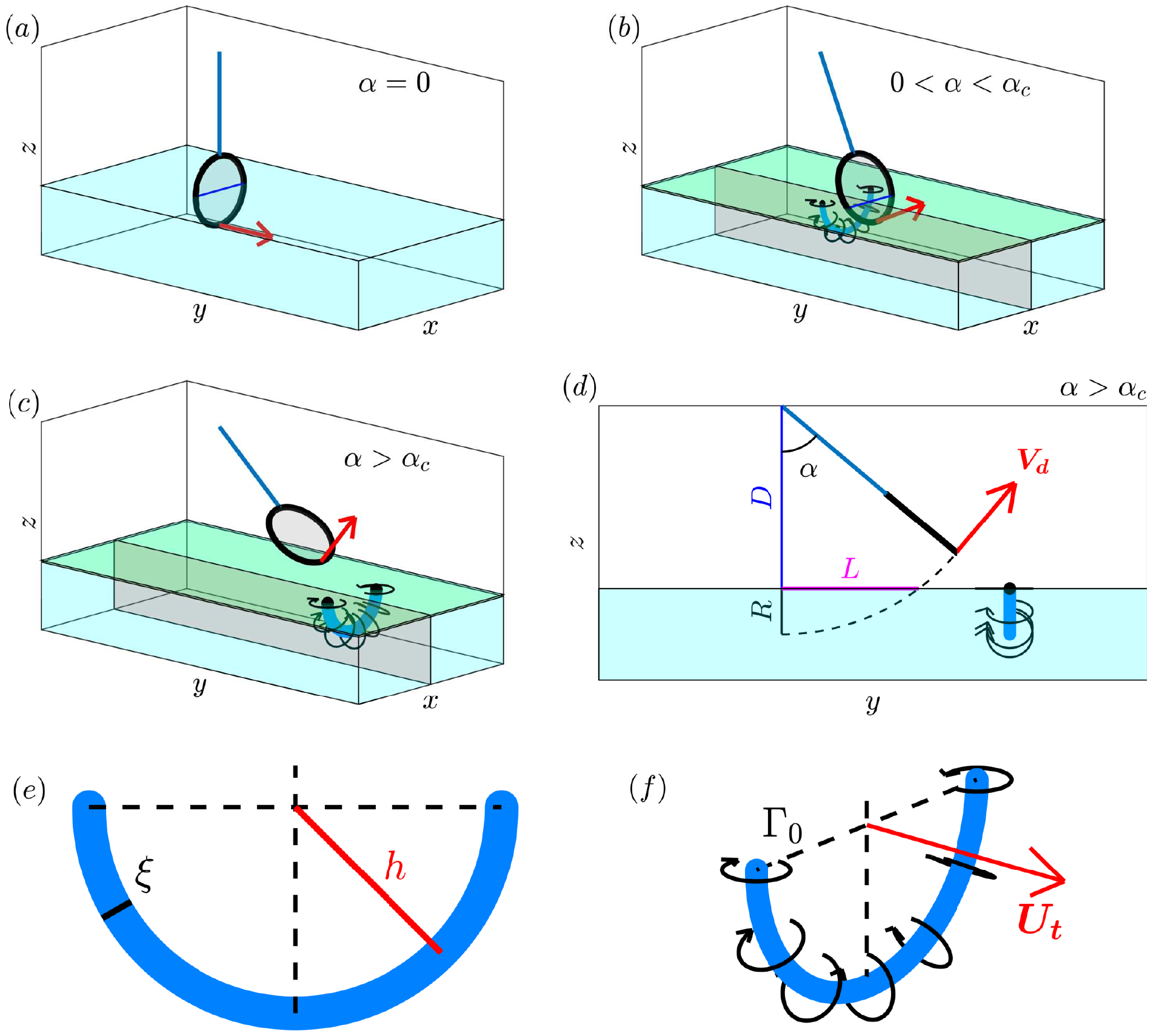}
	\caption{(a) Experimental setup with the disk half-immersed at its initial position. (b) Formation of the semi-vortex ring as long as the disk is partially immersed for $0<\alpha<\alpha_c$ with $\alpha_c=\arctan\left(L/D\right)$. The disk moves with a edge velocity $V_d$ The transparent green and grey planes indicate laser sheet positions used for DPIV measurements. (c) Propagation of the SVR released after that the disk was out of the water for $\alpha>\alpha_c$. (d) Side view of SVR released shown in (c) with $\alpha$ angle definition. Front view (e) and 3d-view (f) of a semi-vortex ring with the definition of the ring radius $h$, displacement velocity $U_t$, core radius $\xi$ and circulation $\Gamma_0$.}
	\label{fig:FigSetup}
\end{figure}

Vortex rings can also be produced by a disk motion in a liquid. Only a few studies concern vortex rings generated by a thin circular disk. Taylor \cite{Taylor1953} predicted the properties of a complete vortex ring generated by a disk with a radius $R$ and a linear displacement velocity $V_d$, by using the relation given by Lamb \cite{Lamb1932}:
\begin{equation}
	U_t=\frac{\Gamma_0}{4\pi h}\left(\log\frac{8h}{\xi}-\frac{1}{4}\right)
	\label{UtLamb}
\end{equation}
$U_t$ and $h$ are respectively the translation velocity and the radius of the vortex ring. $\Gamma_0$ and $\xi$ are the circulation and the radius core of the vortex ring (see draw Figure \ref{fig:FigSetup}). From equation \eqref{UtLamb}, Taylor gave the following expressions to relate the disk radius $R$ and velocity $V_d$ to the vortex ring properties:
\begin{align}
	\frac{h}{R} & =0.816 & \frac{U_t}{V_d} & =0.436 & \frac{\xi}{R} & = 0.152 & \frac{\Gamma_0}{R V_d} & =\frac{4}{\pi}
	\label{TaylorEq}
\end{align}
Sallet \cite{Sallet1975} showed experimentally that the properties of vortex rings generated by a piston in air are well described by these relations. Yang {et al.} \cite{Yang2012} studied experimentally the formation and evolution of a vortex ring generated in water by the linear displacement of a thin circular disk. Similarly to Gharib \emph{et al.} \cite{Gharib1998}, they found a characteristic time formation, a Gaussian-like vorticity distribution in the core and a dimensionless energy value close to that found for vortex rings created by a piston stroke. 

\medskip

In this work, we focus on the properties of a semi-vortex ring (SVR) induced by the circular motion of a flat disk integrated with the end of a rigid pendulum as shown in Figure \ref{fig:FigSetup}.
This vortex is a semi-torus (or U-shaped vortex) formed underwater whose the two ends are connected to the free surface. In the experiments conducted by Yang \emph{et al.} \cite{Yang2012}, the vortex rings are generated by a rolling-up process at the rear of the disk and disappear if the disk stops. In our experiments, a SVR is induced by the circular disk motion and then still propagates, while being connected to the free surface even after the disk was out of the water. While a vortex ring is released from a piston by a pinch-off phenomenon \cite{OFarrell2014}, a SVR is released from the disk and can be considered as a propagating isolated object.

A SVR is connected to a free surface. Bernal and Kwon \cite{Bernal1989} show that an underwater vortex ring moving parallel to a free surface can interact with the latter. It can transfer vorticity to the free surface, or open up and reconnect to this surface. Gharib and Weigand \cite{Gharib1996} studied this reconnection phenomenon for oblique approaching vortex rings. They show that surface properties play a major role for vortex reconnection and disconnection, as well as an eventual generation of a secondary vortex. A SVR produced in our experiments is connected to the free surface from its initial formation. An important issue is the influence of the free surface on the properties of the connected SVR during propagation. In this work, we observe that the SVR circulation decreases without apparent growth of the core radius as expected if the decrease is due to viscous effect. We assume that a vorticity transfer from the SVR to the free surface might be the cause.

Despite this connection to the free surface, the formation process for a SVR is very similar to the formation of vortex rings produced by the caudal fin of some fish \cite{Lauder2002} and SVRs produced by water strider \cite{Dickinson2003}. Indeed SVRs have been already observed by Hu \emph{et al.} \cite{Hu2003} by analysing water strider locomotion. This structure has made it possible to solve the Denny's paradox \cite{Denny1993}: Water strider locomotion was first explained by capillary waves induced by their leg motion \cite{Sun2001} but this assumption could not be valid for water strider infants. High speed videos and particle tracking experiments allowed Hu \emph{et al.} \cite{Hu2003} to observe U-shape vortices connected to the water free surface, explaining the origin of water strider locomotion by application of Newton's third law. However, to our knowledge, the formation mechanism by the circular motion of a thin object like a disk, and the link between these object characteristics and the vortex properties remain open questions. We show here that, in addition to the separation and the rolling-up of the boundary layer in the near wake of a thin disk, an instability called 'transition waves' identified by Bloor \cite{Bloor1964}, occurs and produces secondary SVRs similarly to secondary vortices produced in a cylinder wake \cite{Wei1986}. These secondary vortices modify the final circulation value of the main SVR, which consequently depends of the disk thickness.

\medskip

In order to describe in detail our observations, the rest of the paper is structured as follows. Section \ref{sec:Material} presents the experimental set-up to generate a SVR and the methods used to obtain vortex properties from the observed fluid motions. In Section \ref{sec:Results}, first we describe SVR properties such as dimensions, circulation and dynamics thus we explain how these properties are related to the disk characteristics and what are the effects of the free surface. Then we discuss the particular role of the disk thickness $e$ during the formation process and we highlight the existence of a shedding phenomenon producing secondary vortices which the properties are studied.

\section{Material and Methods}
\label{sec:Material}

This Section describes the experimental set-up and the methods used to obtain the SVR properties (ring radius $h$, displacement velocity $U_t$, core radius $\xi$ and circulation $\Gamma_0$ shown in the schema in Figure \ref{fig:FigSetup}e and \ref{fig:FigSetup}f).

\subsection{Experimental set-up}
\label{sec:Setup}
The experiments are realized in a glass tank of $150\times 60\times 60$ cm filled with water at a height of 30 cm. A thin plexiglas disk (radius $R$ and thickness $e$) is attached to a rigid pendulum which the rod length $D$ can be varied. The pendulum is moved by a linear motor, leading to a circular trajectory for the disk at its extremity (see Figure \ref{fig:FigSetup}).
We note $\alpha$ the angle between the horizontal $y$-axis and the pendulum axis. For the initial vertical position of the disk, $\alpha=0$\textdegree. The linear motor has a linear acceleration during 1 second before reaching its final velocity to avoid violent motion in water and surface waves propagation, capable of interacting with vortices \cite{Humbert2017}. We note $V_d$ the final velocity reached by the disk edge.

The disk, initially in a vertical position and half-immersed, travels a distance $L$ underwater before coming out from water (Figure \ref{fig:FigSetup}) and releases the SVR. The output properties ($h$, $U_t$, $\xi$, $\Gamma_0$) of the SVR can be expressed as a function of $n=4$ input parameters ($R$, $L$, $V_d$, $\nu$), where $\nu$ is the water kinematic viscosity. The particular effect of the disk thickness $e$ will be discussed in the Section \ref{sec:SVRformation}. The input parameters depend on $k=2$ dimensions (length and time). From Buckingham $\pi$ theorem \cite{Rayleigh1892}, the measured output properties of SVR will depend of $n-k=2$ dimensionless quantities. Consequently ($h$, $U_t$, $\xi$, $\Gamma_0$) can be expressed as a function of the Reynolds number associated to the disk motion $Re=RV_d/\nu$, and the dimensionless ratio $L/R$ (see Figure \ref{fig:FigSetup}d). For vortex rings generated by a $R$-radius piston moved over a distance $L$, Gharib {et al.} \cite{Gharib1998} have shown that the ratio $L/R$ has a great influence on formation process. Only for large ratios $L/R>4$, the leading vortex ring generated is followed by a trailing jet. To avoid the formation of this trailing jet, we choose a small constant ratio $L/R=3$, which means that the disk moves underwater from the position $y=0$ to $y/R=3$. Thus in the experiments conducted in Section \ref{sec:SVRproperties}, we only vary the Reynolds number $Re$ by using several disk radii $R$ and disk velocities $V_d$. In order to compare the different experiments, we chose a maximum disk velocity $V_d^{\mathrm{max}}$ such as the product $RV_d^{\mathrm{max}}$ remains constant for all disk radii $R$. Here $RV_d^{\mathrm{max}}/\nu=1.8\times 10^4$ in all experiments and $V_d^{\mathrm{max}}$ is the maximum disk velocity. We note the dimensionless input velocity $v_m^{}=V_d^{}/V_d^{\mathrm{max}}$ for simplicity.

\subsection{DPIV measurements}
\label{sec:DPIV}
\begin{figure}\centering
	\includegraphics[width=12cm,trim=0 0 0 0,clip=true]{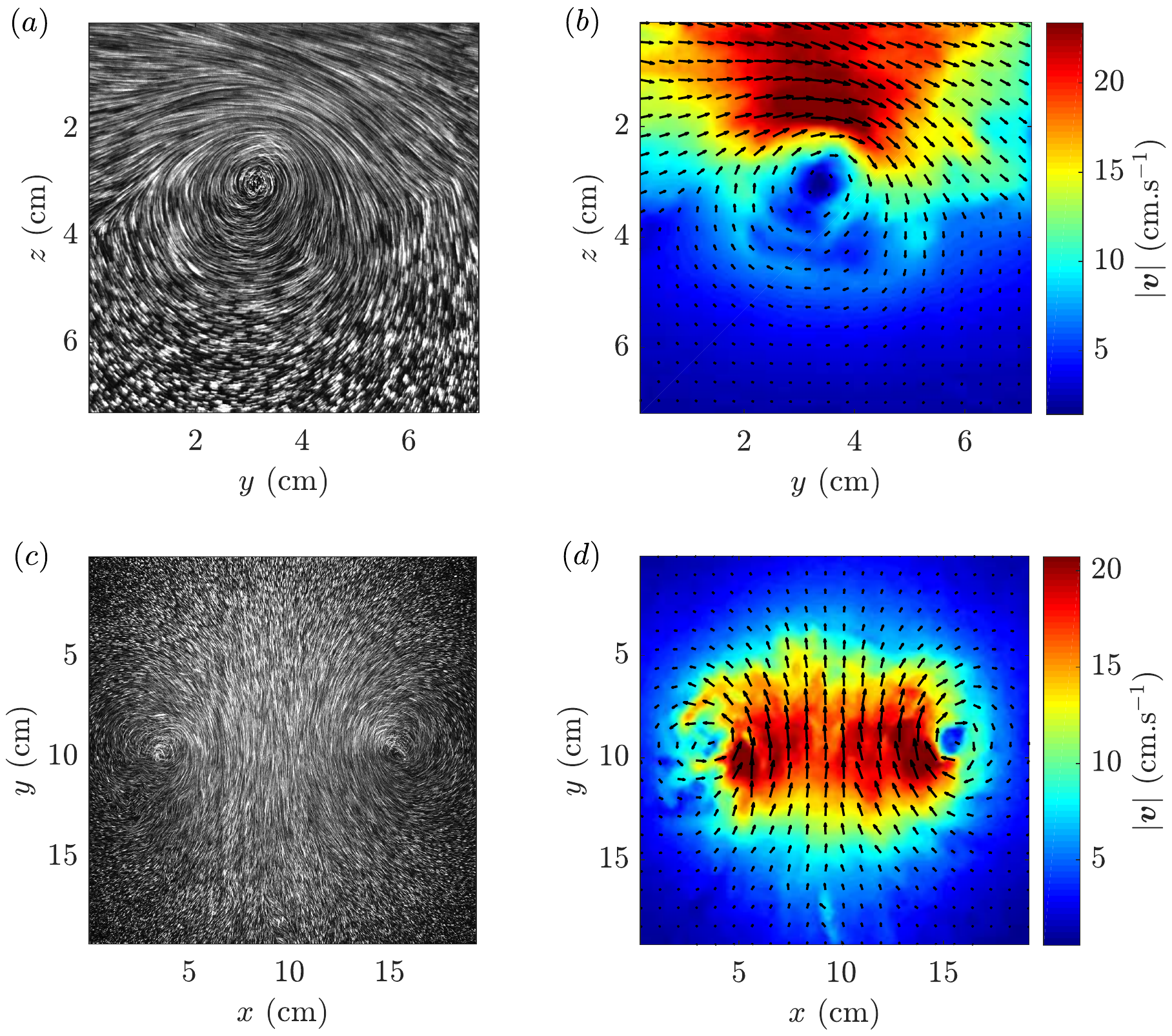}
	\caption{Temporal stack of 25 frames filmed at 500 fps in the vertical plane $0yz$ (a), and in the horizontal plane $0xy$ (c) slightly below the water surface. Velocity fields obtained from PIVlab in the vertical plane $0yz$ (b) and in the horizontal plane $0xy$ (d). Disk parameters: $R=6$ cm, $e=4$ mm, $V_d=15$ cm.s$^{-1}$. The frame field centre is at the position $y=45.5$ cm from the pendulum axis.}
	\label{fig:FigurePIV}
\end{figure}

Velocity fields are obtained by Digital Particle Image Velocimetry. The water is seeded with particles of 20 $\mu$m diameter with a density close to water density and the tank is illuminated by a laser sheet (wavelength: 532 nm). A high speed camera, perpendicular to the laser sheet, is used to record the particle motion, and subsequently to determine the velocity fields \cite{Raffel2007}. In our experiments, we focus on two particular planes: (i) the vertical plane $0yz$ (grey plane in Figure \ref{fig:FigSetup}), a side view to observe the bottom part of the SVR and (ii) the horizontal plane $0xy$ (green plane in Figure \ref{fig:FigSetup}), a bottom view to obtain the structure of the SVR just below the surface. The former shows the propagation of a unique vortex along $y$-axis (Figure \ref{fig:FigurePIV}a and \ref{fig:FigurePIV}b) while the latter shows two counter-rotating vortices (Figure \ref{fig:FigurePIV}c and \ref{fig:FigurePIV}d).

The videos are recorded with a frame rate of 500 frames per second (fps) and a resolution of $1024 \times 1024$ pixels. The DPIV analysis is performed with the Matlab code PIVlab \cite{Thielicke2014} with an interrogation area of $32 \times 32$ pixels for single vortex views and $16 \times 16$ pixels for double vortex views. Few videos were performed at different distances $y$ from the pendulum axis, allowing to follow the SVR during its propagation with a sufficient spatial resolution and to analyse core properties by DPIV measurements. Figure \ref{fig:FigurePIV} shows a 25-frame stack and the corresponding velocity fields. The temporal averaging reveals the streamlines and then the vortex structure, in agreement with the velocity fields obtained in both cases.

\section{Results}
\label{sec:Results}

First we present measurements about the shape and core structure of the SVR in Section \ref{sec:SVRstructure}. Then in Section \ref{sec:SVRproperties}, the properties of the SVR are studied (ring radius $h$, displacement velocity $U_t$, core radius $\xi$, circulation $\Gamma_0$), and in particular, their dependence as function of the disk characteristics (radius $R$, edge velocity $V_d$, thickness $e$). The secondary vortices, produced by the shedding phenomenon due to the finite disk thickness $e$, are finally discussed in the Section \ref{sec:SVRformation}.

\subsection{Structure of a semi vortex ring}
\label{sec:SVRstructure}

\begin{figure}\centering
	\includegraphics[width=12cm,trim=0 0 0 0,clip=true]{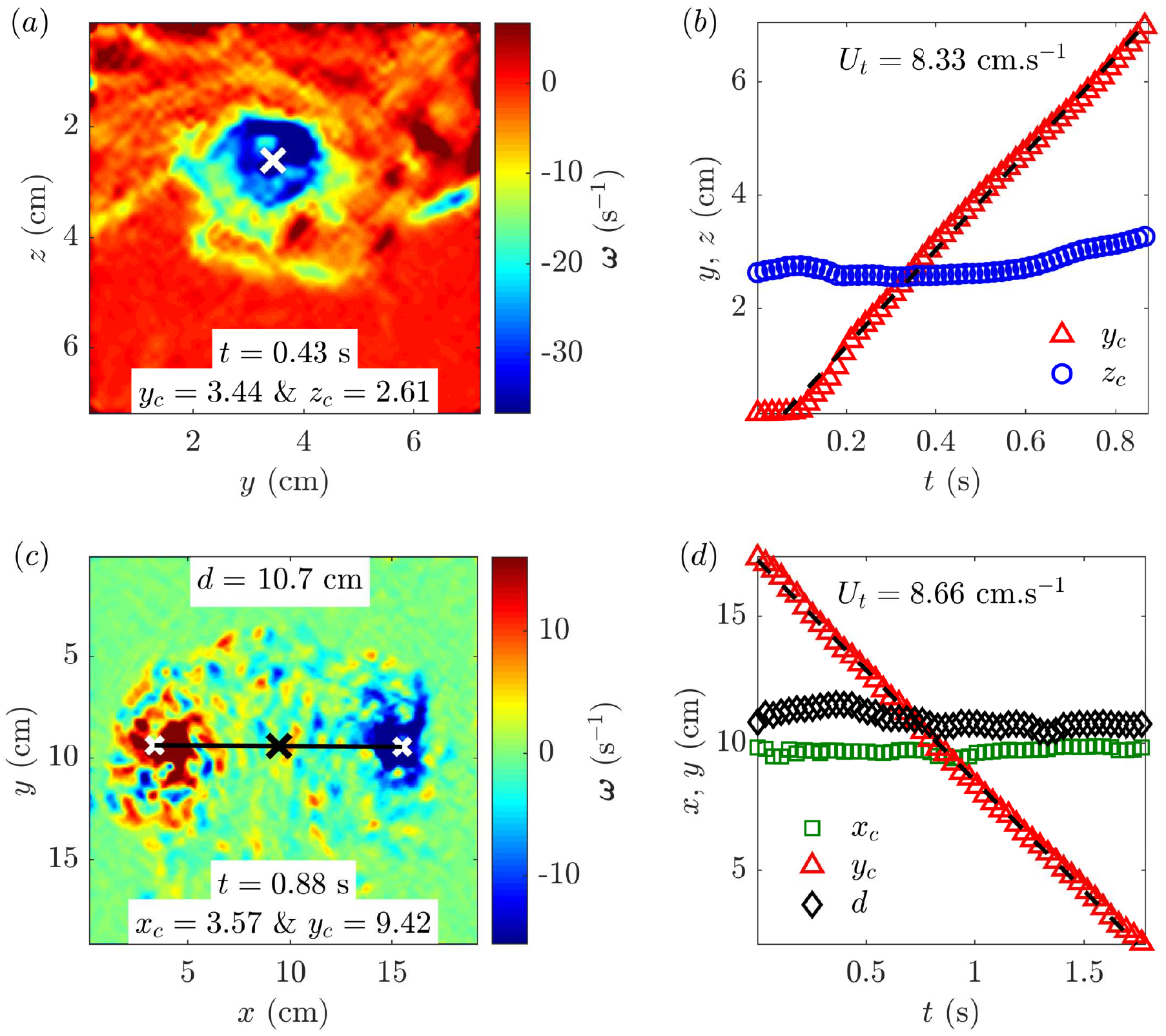}
	\caption{Examples of vorticity fields $\boldsymbol{\omega}$ obtained in the plane $0yz$ (a) and $0xy$ (c). Displacement of the vortex centres in the vertical plane $0yz$ (b), and in the horizontal plane $0xy$ (d) below the water surface. The blue circles and the red triangles are experimental measurements, the black dashed line is a linear-regression model to obtain the displacement velocity $U_t=\partial_t y_c$. Disk parameters: $R=6$ cm, $e=4$ mm, $V_d=15$ cm.s$^{-1}$. The frame field centre is at the position $y=45.5$ cm from the pendulum axis.}
	\label{fig:FigureRotCenter}
\end{figure}

To describe the vortex properties, we first accurately localized the vortex centres from the vorticity field $\boldsymbol{\omega}=\boldsymbol{\nabla}\wedge\boldsymbol{v}$, obtained from velocity fields in both planes (Figures \ref{fig:FigureRotCenter}a and \ref{fig:FigureRotCenter}c).
A centroid detection on $\left|\boldsymbol{\omega}\right|$ provides the vortex displacement. In the vertical plane $0yz$, ($y_c ; z_c$) are the coordinates of the centre of the bottom part of the SVR. In the horizontal plane $0xy$, ($x_c ; y_c$) are the coordinates of the barycentre of the two extremities separated by a distance $d$. The displacement velocity is defined in both cases by $U_t=\partial_t y_c$ while $x_c$ and $z_c$ remain constant. Figures \ref{fig:FigureRotCenter}d and \ref{fig:FigureRotCenter}d show that the SVR has a quasi-constant velocity displacement $U_t$ for a distance of a few centimetres. Although displacement velocity of vortex rings decreases due to viscous dissipation \cite{Weigand1997}, we are not able to measure this variation for a short distance. We will see further in this paper that the decrease of SVR displacement velocity can be measured for larger distances. The constant values obtained for position $z_c$ of the SVR bottom part and for the distance $d$ between the two SVR extremities, provides informations about SVR ring radius $h$. They also prove that the shape of the SVR is circular as discussed below.

\begin{figure}\centering
	\includegraphics[width=12cm,trim=0 0 0 0,clip=true]{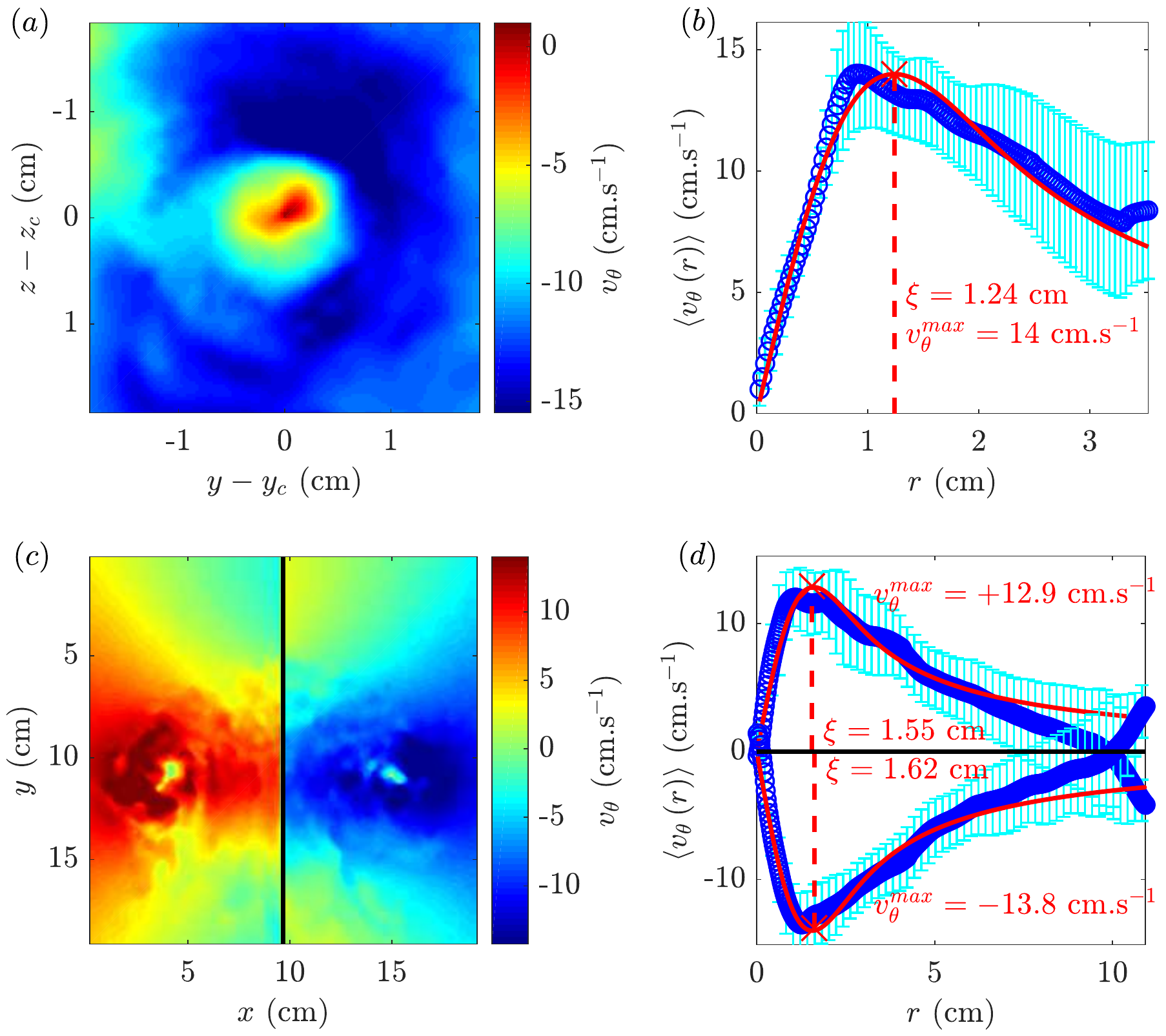}
	\caption{Examples of azimuthal fields $v_{\theta}$ obtained in the planes $0yz$ (a) and $0xy$ (c). Azimuthal velocity profiles $\left<v_{\theta}\right>$, angularly averaged in the vertical plane $0yz$ (b), and in the horizontal plane $0xy$ (d) slightly below the water surface. The blue circles are experimental measurements and the red lines represent the Lamb-Oseen model \cite{Saffman1992} used to determined the core properties $\xi$ and $\Gamma_0$. Disk parameters: $R=6$ cm, $e=4$ mm, $V_d=15$ cm.s$^{-1}$. The frame field centre is at the position $y=45.5$ cm from the pendulum axis.}
	\label{fig:FigureVtheta}
\end{figure}

Properties of the vortex core are hereafter studied in the reference frame of the vortex centres. For each video, azimuthal velocity fields $v_{\theta}$ are calculated after subtracting the displacement velocity $U_t$ (Figures \ref{fig:FigureVtheta}a and \ref{fig:FigureVtheta}c). 
In Figures \ref{fig:FigureVtheta}b and \ref{fig:FigureVtheta}d, we observe the vortex core characteristics in the vertical plane $0yz$. The azimuthal velocity field is angularly averaged to obtain the velocity profile $v_{\theta}\left(r\right)$, where $r$ is the distance form the vortex axis (Figures \ref{fig:FigureVtheta}a and \ref{fig:FigureVtheta}c). The velocity profile $v_{\theta}\left(r\right)$ can be decomposed in two regions: the vortex core where the velocity increases from zero to $v_{\theta}\left(r=\xi\right)=v_{\theta}^{\mathrm{max}}$ for $r<\xi$ and a second irrotational region where the velocity decreases for $r>\xi$. This type of velocity profile can be captured by Lamb-Oseen model \cite{Saffman1992} that assumes a Gaussian vorticity distribution for the vortex core, also verified for vortex rings generated by a piston stroke \cite{Weigand1997}:
\begin{equation}
	v_{\theta}\left(r\right)=\frac{\Gamma_0}{2\pi r}\left[1-\exp\left(-\left(\frac{r}{\xi}\right)^2\right)\right]
	\label{LambOseen}
\end{equation}
Where $\Gamma_0$ is the circulation far from the vortex axis, assumed to be constant in this model. Fitting our experiments by this model, we are able to obtain the core circulation $\Gamma_0$ and the core radius $\xi$ (Figures \ref{fig:FigureVtheta}b and \ref{fig:FigureVtheta}d). The SVR properties do not evolve for the duration of a given experiment ($\sim 1.4$ s). As a consequence, thereafter we note $\xi=\left<\xi\right>$ and $\Gamma_0=\left<\Gamma_0\right>$, where $\left<A\right>$ is the temporal mean value of a quantity $A$ for a given experiment.

\begin{figure}\centering
	\includegraphics[width=12cm,trim=0 0 0 0,clip=true]{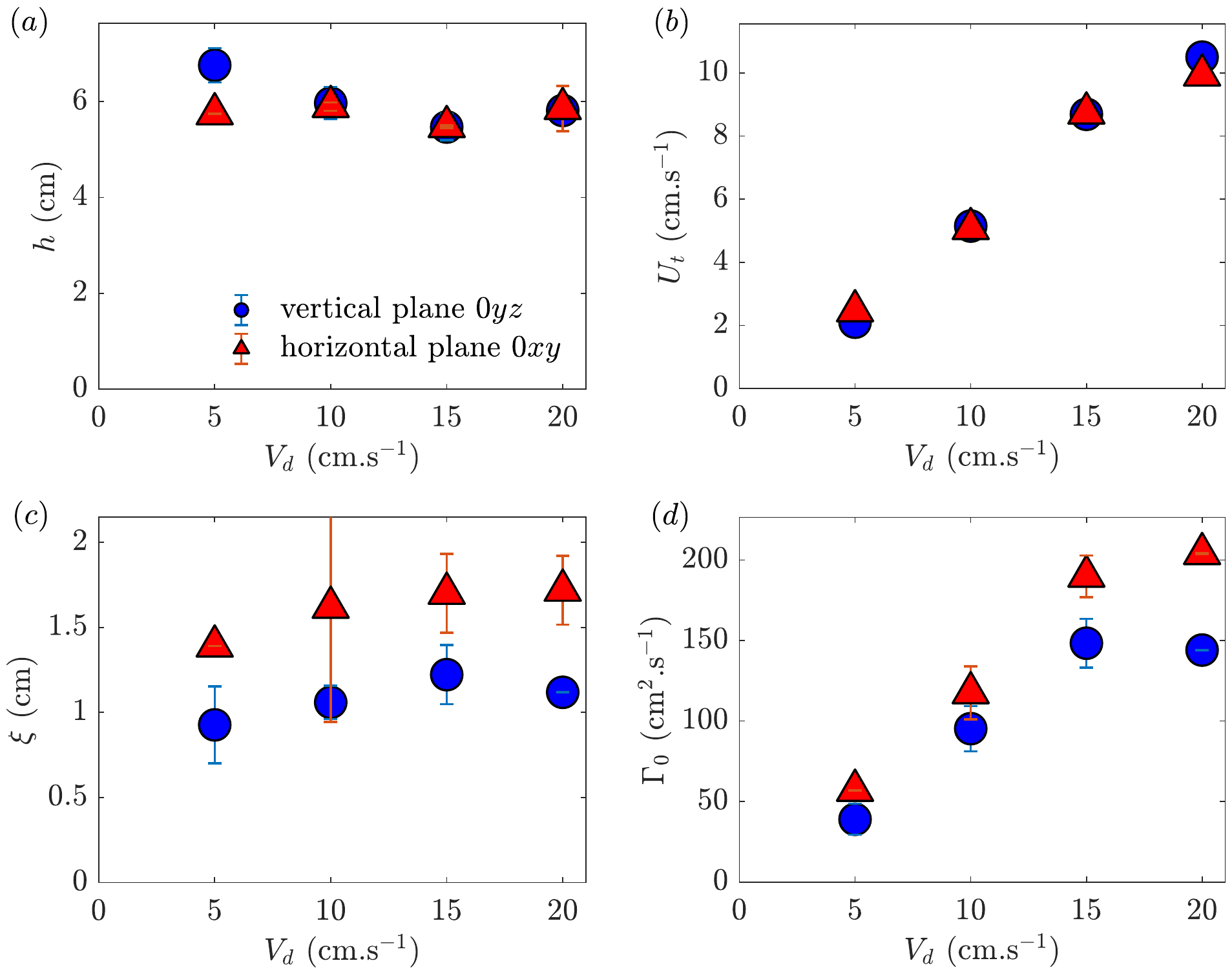}
	\caption{Measurements of ring radius $h$ (a), displacement velocity $U_t$ (b), core radius $\xi$ (c) and circulation $\Gamma_0$ (d) in the  horizontal plane $0yz$ (side view) and in the vertical plane $0xy$ (bottom view). Disk parameters: $R=6$ cm, $e=4$ mm. The frame field centre is at the position $y=45.5$ cm from the pendulum axis.}
	\label{fig:Fig2plans}
\end{figure}

We carried out measurements of SVR properties (ring radius $h$, displacement velocity $U_t$, core radius $\xi$ and circulation $\Gamma_0$) versus disk velocities $V_d$ with $R=6$ cm and $e=4$ mm at a distance $y=45.5$ cm from the pendulum axis in the vertical and horizontal planes (Figure \ref{fig:Fig2plans}). In the vertical plane $0yz$, the ring radius $h$ (see Figure \ref{fig:FigSetup}f) is defined by $h=\left<z_c\right>+h_0$, where $\left<z_c\right>$ is the temporal mean value and $h_0$ is the water depth from the free surface to the frame centre (see Figure \ref{fig:FigureRotCenter}b). In the horizontal plane $0xy$, the ring radius is defined by $h=\left<d\right>/2$, where $d$ is the distance between the two extremities (see Figure \ref{fig:FigureRotCenter}d).
For all quantities, we find close values for the bottom part of the SVR and its extremities just below the surface. These similar values for $h$, $\xi$ and $\Gamma_0$ in both plans emphasize that the SVR is a tube or constant vorticity, semi-circle shaped. In addition, the similar $U_t$ in the bottom part of the SVR and below the surface, show that this semi-circle propagates without stretching or deformation.

Furthermore, Figure \ref{fig:Fig2plans} shows that the ring and core radius do not depend on the disk velocity whereas the displacement velocity and circulation increase with the disk velocity, in agreement with Taylor predictions \eqref{TaylorEq}. In the next section, we turn attention to the dependence of the vortex properties on the disk characteristics.

\subsection{Dynamics of the SVR properties and influence of the disk characteristics}
\label{sec:SVRproperties}

We studied the evolution of SVR properties for several distances $y$ from the pendulum axis ($y=0$ cm corresponds to the initial vertical position of the half-immersed disk). Measurements have been carried out for three disks of different radii ($R=6$, 8, 10 cm) and a constant thickness $e=4$ mm. We vary the Reynolds number $Re$ associated to the disk motion and maintain constant the dimensionless ratio $L/R$ as discussed in Section \ref{sec:Setup}.

\begin{figure}\centering
	\includegraphics[width=13.5cm,trim=0 0 0 0,clip=true]{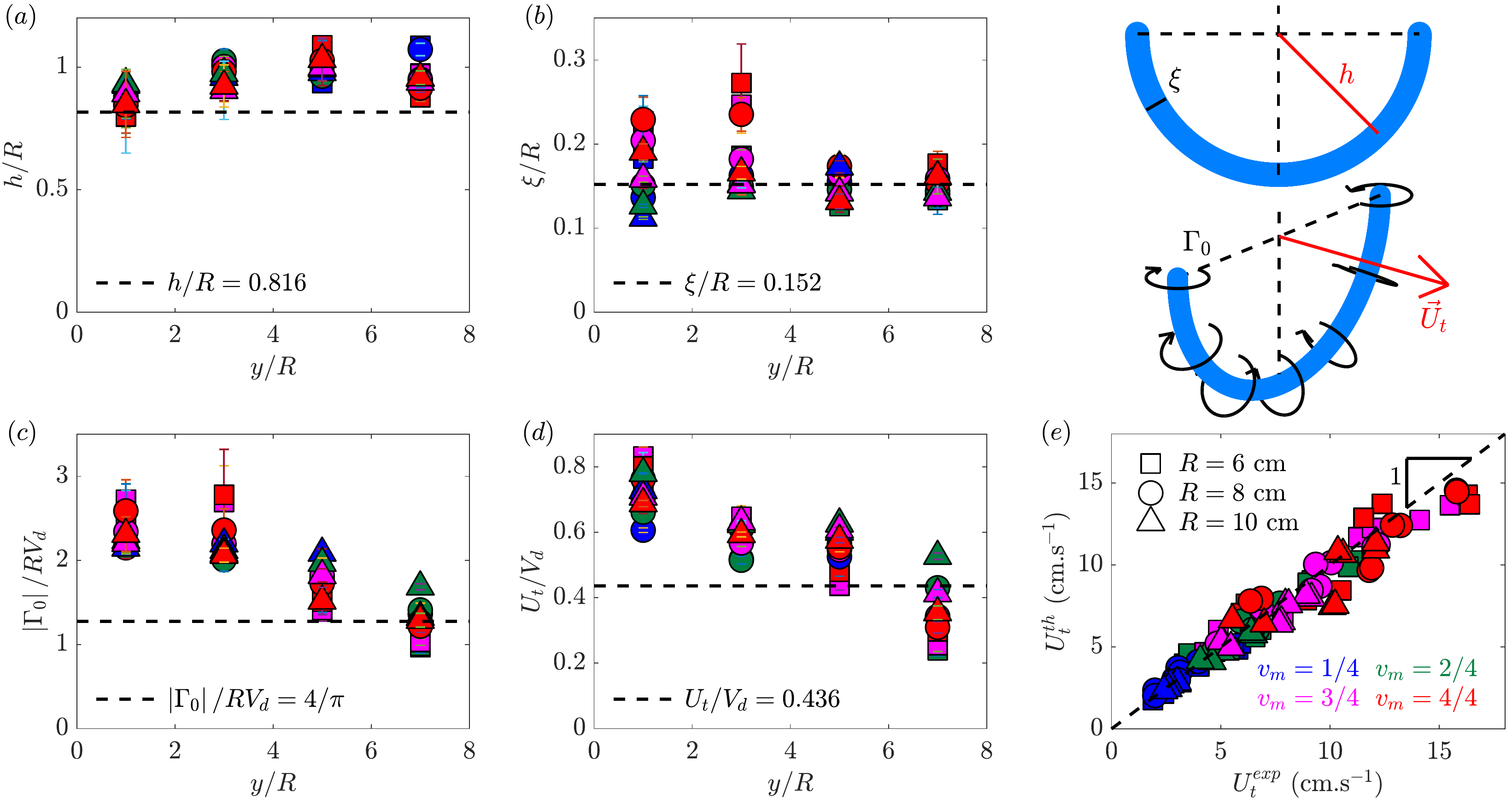}
	\caption{Measurements of ring radius $h$ (a), the core radius $\xi$ (b), the circulation $\Gamma_0$ (c) and the velocity displacement $U_t$ (d) for different distances $y$ from the pendulum axis. The black dashed line indicates the dimensionless values given by equations \eqref{TaylorEq} predicted by Taylor \cite{Taylor1953}. (e) Theoretic displacement velocity $U_t^{th}$ given by equation \eqref{UtLamb} predicted by \cite{Lamb1932} versus the experimental measurements $U_t^{exp}$ shown in Figure \ref{fig:FigNormPropRV}d. The black dashed line has a slope $a=1$. Parameters: $L/R=3$, $e=4$ mm, $RV_d^{\mathrm{max}}/\nu=1.8\times 10^4$ and $v_m^{}=V_d^{}/V_d^{\mathrm{max}}$.}
	\label{fig:FigNormPropRV}
\end{figure}

Figure \ref{fig:FigNormPropRV} shows the ring radius $h$ (a), the core radius $\xi$ (b), the circulation $\Gamma_0$ (c) and the velocity displacement $U_t$ (d) normalized by the disk characteristics ($R$, $V_d$) versus $y$-distance from pendulum axis.

First, we observed that the ring and core radius have constant values during propagation and can be rescaled by the disk radius $R$. This result can be surprising as the core radius $\xi$ is assumed to grow by viscous diffusion as $\xi\left(t\right)=\xi_0+\sqrt{\nu t}$ where $\xi_0$ is the initial core radius at the end of the formation process \cite{Saffman1970}. Let us consider the slowest SVR travelling the largest distance in our experiments. After its release from a disk $R=10$ cm, this SVR propagates at $U_t\sim 3$ cm.s$^{-1}$ over 40 cm, during 13 s. During this time, the core size variation can be established $\delta\xi\sim 3.6$ mm which means $\delta\xi/\xi\sim 20\%$, which is smaller than the uncertainties on $\xi$. Thus, we assume that, though the core radius increases during propagation, our measurements are not accurate enough to observe such temporal variation over a distance of $4R$. Despite a different geometry, the dimensionless quantities $h/R$ and $\xi/R$ are in a good agreement with relations \eqref{TaylorEq} predicted by Taylor \cite{Taylor1953} for a full vortex ring generated by the linear displacement of a flat disk.

Figure \ref{fig:FigNormPropRV}c shows that the circulation can be well normalized by the product $RV_d$, as predicted by the Buckingham $\pi$ theorem for a constant disk thickness $e=4$ mm and a constant ratio $L/R=3$. The dimensionless circulation decreases significantly after $y/R=3$, corresponding to the moment when the disk comes out of the water and energy injection to the liquid ends. The circulation decay has been already observed for vortex rings generated by a piston stroke \cite{Weigand1997}. However, in this previous case, the circulation and velocity decays where induced by the core expansion due to viscosity. As we do not observe such expansion, maybe due to a lack of resolution, we assume that the observed decrease in our setup, is not only due to the viscous dissipation but also and probably mainly to a vorticity transfer from SVR to the free surface \cite{Bernal1989}.

It is noteworthy that the initial circulation value of the SVR for $y/R\leq 3$ is larger than the dimensionless value $\Gamma_0/RV_d=4/\pi$ predicted by Taylor \cite{Taylor1953}. For vortex rings generated by a piston stroke, Didden \cite{Didden1979} calculated the initial circulation value by demonstrating that the vorticity flux is given by $U_t^2/2$ for a constant linear displacement velocity $U_t$ of the piston. Because of the initial acceleration and final deceleration of a real piston, Glezer \cite{Glezer1988} showed that the calculation of this initial circulation is more complex and depends on the complete history of the piston displacement. Predicting the initial circulation in our experiments requires to obtain the vorticity flux for the circular motion of a disk. We will see in Section \ref{sec:SVRformation} that the vorticity flux also depends on the disk thickness $e$ for a SVR generated by the circular motion of a thin disk.

Figure \ref{fig:FigNormPropRV}d shows that the displacement velocity $U_t$ can be rescaled by the disk edge velocity $V_d$. Similarly to the circulation, $U_t$ is larger than the value predicted by Taylor \cite{Taylor1953} and decreases during propagation (Figure \ref{fig:FigNormPropRV}d). According to equation \eqref{UtLamb} and because the ring radius $h$ and core radius $\xi$ remain constant during propagation, we can here approximate $U_t \propto \Gamma_0$, which is also in agreement with the similarity of velocity displacement and circulation decay. From the experimental measurements ($h$, $\xi$, $\Gamma_0$), we calculated the theoretical displacement velocities $U_t^{th}$ given by equation \eqref{UtLamb} as shown in Figure \ref{fig:FigNormPropRV}e. Although we have a semi-vortex ring connected to a free surface, the vortex properties are in a very good agreement with the prediction given by Lamb \cite{Lamb1932} for a full vortex ring.

\medskip

To summarize, the ring and core radii ($h$ and $\xi$), only determined by the disk radius $R$, remain constant during propagation and obey to Taylor's predictions given by equations \eqref{TaylorEq}. We assume that the decrease in both $U_t$ and $\Gamma_0$ are not only due to the viscous dissipation but also to vorticity transfer from SVR to the free surface. Though our SVR is connected to a free surface, its properties are also described by the Lamb's relation given by equation \eqref{UtLamb}, assumed to be valid for a complete vortex ring. Contrary to ring and core radii, velocity displacement and circulation are larger than Taylor's prediction given by equation \eqref{TaylorEq}. This larger circulation value could be partially due to our particular geometry but in the following Section \ref{sec:SVRformation}, we will detail the particular role of  the disk thickness $e$ one the formation process and its influence one the final value of the main SVR circulation $\Gamma_0$.

\subsection{Formation process and secondary vortices}
\label{sec:SVRformation}

\subsubsection{Shedding process}

\begin{figure}\centering
	\includegraphics[width=12cm,trim=0 0 0 0,clip=true]{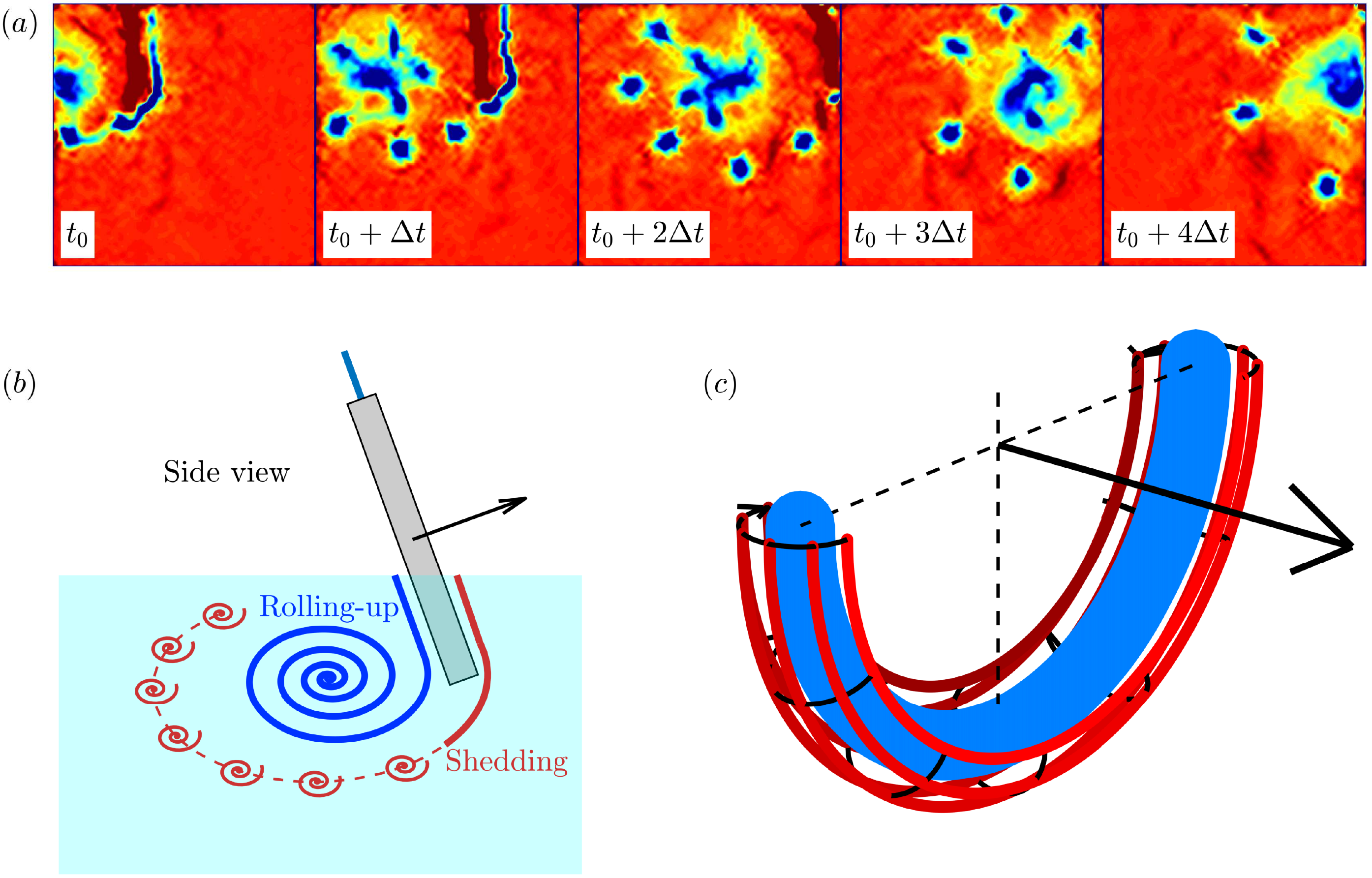}
	\caption{(a) Measurements of the vorticity fields $\boldsymbol{\omega}$ versus time for a disk radius $R=10$ cm, thickness $e=4$ mm and velocity $V_d=14$ cm.s$^{-1}$, $L/R=3$. (b) Side view of the rolling-up on the rear side and shedding phenomena on the leading edge. (c) Semi vortex ring surrounded by secondary vortices produced by the shedding phenomenon.}
	\label{fig:FigSMdraw}
\end{figure}

By recording DPIV images from the initial position of the disk, we are able to analyse the first stage of the SVR formation. As for many configurations where fluid flows past an obstacle \cite{King1977}, the vorticity fields exhibit a shedding phenomenon during formation (Figure \ref{fig:FigSMdraw}a). If the flow velocity is large enough, the boundary layer periodically separates and rolls up into vortices, which are smaller than the main one, due to Kelvin-Helmotz instability (see Figure \ref{fig:FigSMdraw}b). 
For reasons discussed further, we call these smaller vortices `secondary vortices'. Such a shedding phenomenon is clearly observed on the disk edge. Meanwhile, on the rear side of the disk, the boundary layer, formed during disk motion, separates and rolls up into one big SVR. This main SVR is released from the disk, by analogy with the rolling up process observed for vortex rings generated by a piston stroke \cite{Didden1979}.

Considering these observations, we investigate further the interaction between the secondary vortices and the main SVR. On the front face of the disk, the boundary layer separates, is driven toward the rear of the disk and undergoes a shedding phenomenon on the disk edge. Thus the main SVR generated at the rear will attract the secondary vortices as described as described in Figure \ref{fig:FigSMdraw}b and in Movie 1 \cite{SSVRsEVd}. Given that the shedding also occurs in the horizontal plane $0xy$\footnote{Data not shown.}, the secondary vortices are actually smaller SVRs, attracted around the main SVR at the rear of the disk as schematically shown in Figure \ref{fig:FigSMdraw}c. For clarity in the rest of this paper, we call `secondary vortices' the smaller SVRs produced by the shedding phenomenon in contrast to the main SVR, which is bigger and produced by the rolling-up process at the rear of the disk. The secondary vortices are connected to the free surface and attracted toward the rear of the disk to finally merge with the main SVR. We assume that the circular displacement of our thin disk generates such a phenomenon in water, as previously visualized in air for a plate by Pierce \cite{Pierce1961}.

\begin{figure}\centering
	\includegraphics[width=13cm,trim=0 0 0 0,clip=true]{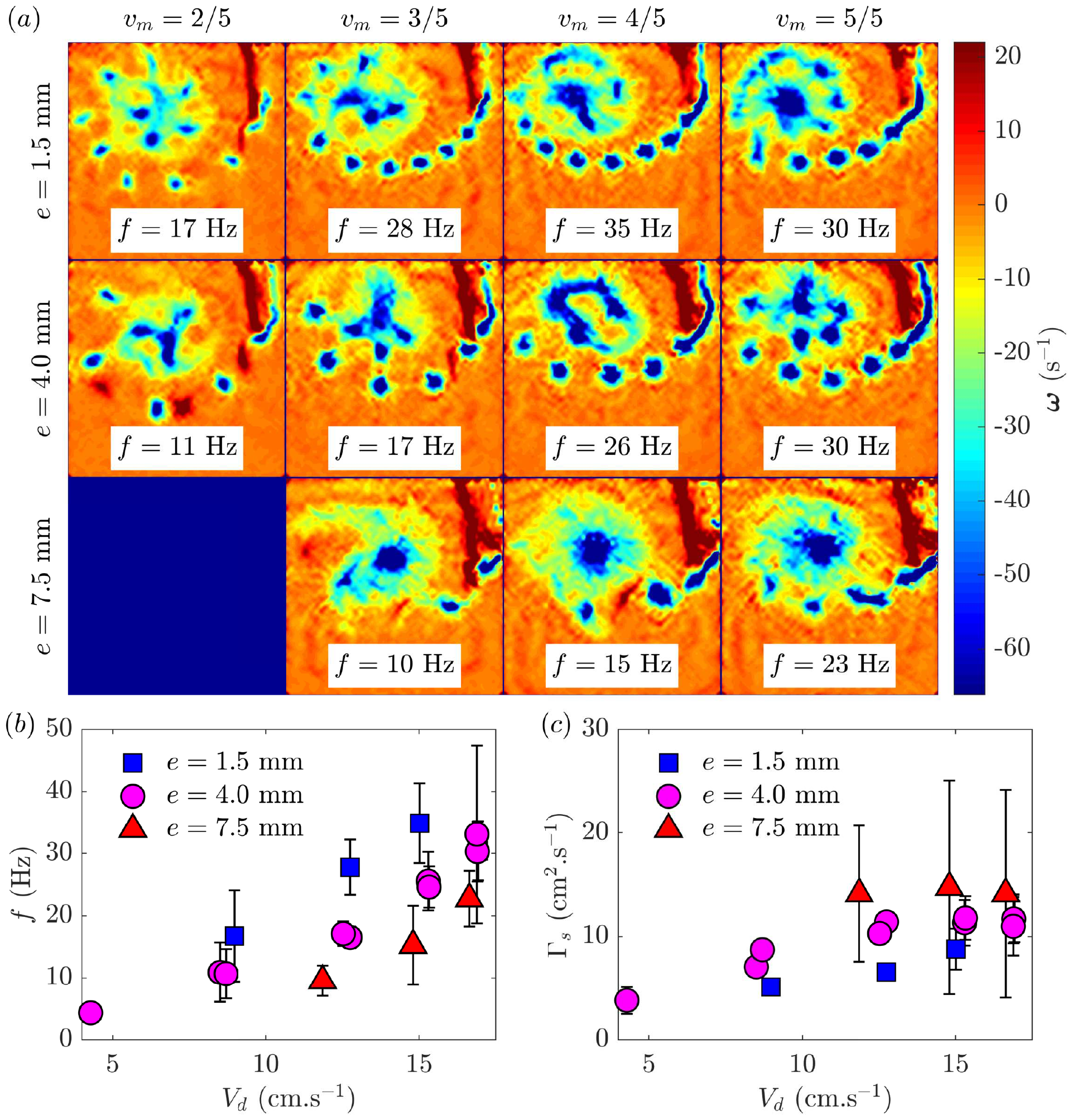}
	\caption{(a) Measurements of the vorticity fields $\boldsymbol{\omega}$ for three disk thickness $e$ and four disk velocities $V_d$ at a distance $y/R=0.68$ from the pendulum axis. Emission frequency $f$ (b) and circulation $\Gamma_s$ (c) of the secondary vortices. The pendulum axis angle is $\alpha=0$\textdegree \ which means the disk had an angular motion of $\Delta\alpha=10$\textdegree \ from its initial vertical position. Parameters: $L/R=3$, $R=10$ cm, $RV_d^{\mathrm{max}}/\nu=1.8\times 10^4$ and $v_m^{}=V_d^{}/V_d^{\mathrm{max}}$.}
	\label{fig:FigSM_VT}
\end{figure}

To understand further the formation process of a SVR and its dependence on characteristic properties of the setup, we carried out vorticity field measurements for three disks with different thickness ($e=1.5$, 4.0, 7.5 mm), a constant radius ($R=10$ cm) and several velocities $V_d$. For these measurements, we focus on the beginning of the formation process for a distance $y/R=0.68$ from the pendulum axis (Figure \ref{fig:FigSM_VT}a).
We observe qualitatively that the emission rate of secondary vortices increases when increasing $V_d$ and decreasing $e$. In order to study this shedding process, we obtain the secondary vortex trajectories by using a centroid detection method. To quantify the emission frequency $f$ related to this process, we measured the times at which the secondary vortices reach a distance of 1.5 cm from the edge of the disk. For a disk motion from the initial vertical position $\alpha=0$\textdegree \ to a position where $\alpha=10$\textdegree, we define the emission frequency by $f=1/\left<\Delta t\right>$, where $\left<\Delta t\right>$ is the mean value of the durations between two small vortex detections. These measurements confirm the observations according to which that the emission frequency $f$ increases with the disk velocity $V_d$  and decreases with the disk thickness $e$ as shown in Figure \ref{fig:FigSM_VT}b and in Movie 1 \cite{SSVRsEVd}.

From the trajectories of the secondary vortices, we also get the circulation of each small vortex. The mean value $\Gamma_s$ of the circulations of secondary vortices for each set of parameters ($e$ ; $V_d$) is plotted in Figure \ref{fig:FigSM_VT}c. We observe that the circulation $\Gamma_s$ increases with increasing disk velocity $V_d$ and also with disk thickness $e$. Thus, the number of secondary vortices produced and their strength increases with increasing disk velocity, in agreement with the dependence of the the final circulation $\Gamma_0$ of the main SVR on $V_d$. The effect of the thickness $e$ is more complex. Indeed the emission frequency of secondary vortices decreases with increasing thickness but the strength of each of them increases. The global effect of the disk thickness on the SVR final circulation $\Gamma_0$ is discussed below.

\subsubsection{Properties of secondary vortices}

We focus now on the properties of the secondary vortices that are the emission frequency $f$ and the circulation $\Gamma_s$.
The shedding phenomenon is characterized by the dimensionless Strouhal number $St=fL/V$ \cite{Strouhal1878}, constant for a given geometry, where $f$, $L$ and $V$ are respectively the vortex emission frequency, a characteristic length of the obstacle and the flow velocity. As shown above, the emission frequency depends on the disk thickness $e$ and velocity $V_d$ so the Strouhal number has to be built with these two quantities. Now we need to detail the instability occurring in the boundary layer on the leading edge of the disk. Indeed the secondary vortices are here produced by a phenomenon called `transition waves' which is a bi-dimensional instability occurring in free shear layers \cite{Bloor1964}. The amplification of this instability may lead to the formation of secondary vortices as observed by Wei and Smith \cite{Wei1986} in the rear-wake of cylinders. For this instability, Ho and Huerre \cite{Ho1984} showed that the appropriate scale for the Stouhal number is the momentum thickness $\theta$ of the boundary layer defined for an incompressible flow by:
\begin{equation}
	\theta=\int\limits _{0}^{\infty}\frac{v\left(y\right)\left(V-v\left(y\right)\right)}{V^{2}}dy
	\label{MomentumThickness}
\end{equation}
Where $V$ is the flow velocity. By assuming a laminar boundary layer, the numerical solution of Blasius equation leads to $\theta\approx0.665\sqrt{\nu e/V_d}$. Finally the Strouhal number is here defined by:
\begin{equation}
	St_{\theta}=\frac{f\theta}{V_d}\approx 0.665f\nu^{1/2}e^{1/2}V_d^{-3/2}
	\label{Strouhal}
\end{equation}
Figure \ref{fig:FigDiscWidth}a shows the value of the Strouhal number $St_{\theta}$ obtained in our experiments with the three different thickness but also the previous experiments with three different radii presented in Section \ref{sec:SVRproperties}.
\begin{figure}\centering
	\includegraphics[width=13cm,trim=0 0 0 0,clip=true]{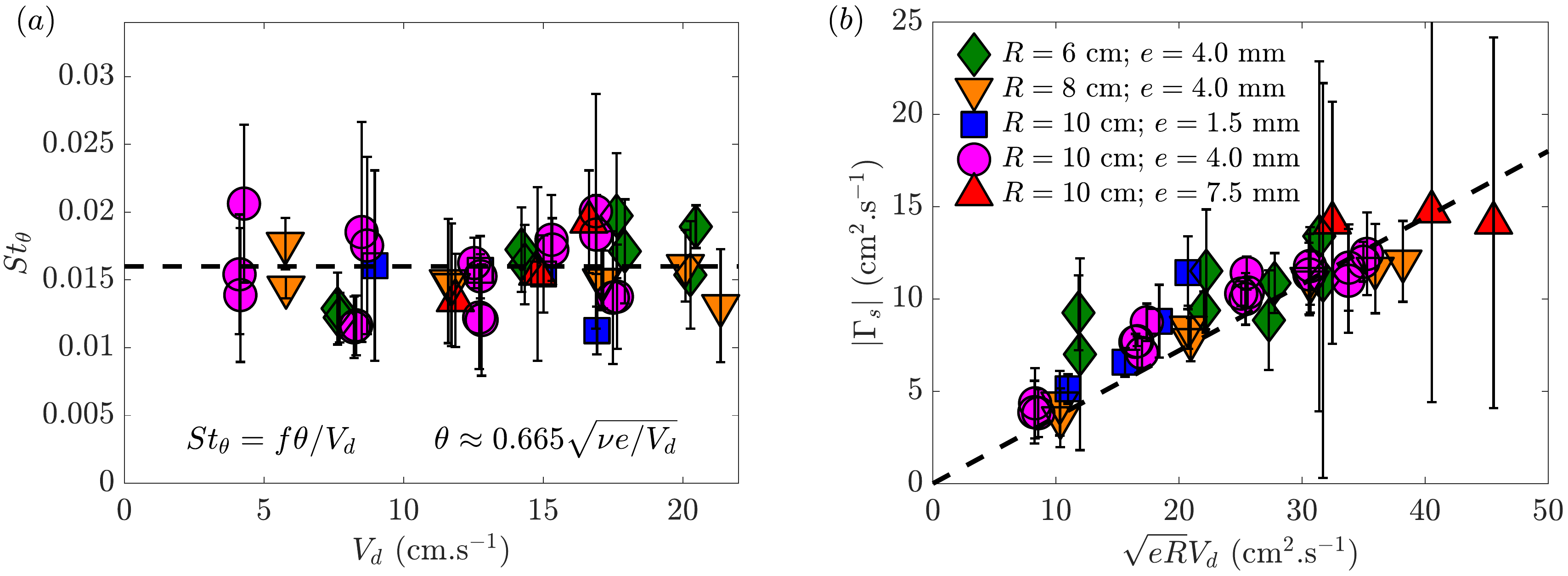}
	\caption{(a) Strouhal numbers $St$ defined by equation \eqref{Strouhal} measured for a disk angular motion $\Delta \alpha=10$\textdegree \ at a distance $y/R=0.68$ from the pendulum axis. The black dashed lines indicate the value $St_{\theta}\approx 0.016$ corresponding to the most amplified wave \cite{Ho1984}. (b) Circulation mean values $\Gamma_s$ of the secondary vortices versus the disk characteristics. Parameters: $L/R=3$, $R=10$ cm, $RV_d^{\mathrm{max}}/\nu=1.8\times 10^4$ and $v_m^{}=V_d^{}/V_d^{\mathrm{max}}$. Dimensionless circulations measured at a distance $y/R=3.18$ from the pendulum axis versus emission frequency $f$. The black dashed line indicates the dimensionless value $\Gamma_0/RV_d=4/\pi$ predicted by Taylor \cite{Taylor1953}. Parameters: $L/R=3$, $R=10$ cm, $RV_d^{\mathrm{max}}/\nu=1.8\times 10^4$ and $v_m^{}=V_d^{}/V_d^{\mathrm{max}}$.}
	\label{fig:FigDiscWidth}
\end{figure}
For all these experiments, the Strouhal number is close to 0.016, which corresponds to the value of the most amplified wave \cite{Ho1984}. This result confirms that the formation of the secondary vortices is due to the bi-dimensional instability identified by Bloor \cite{Bloor1964}. This mechanism is different than the classical shedding phenomenon producing Strouhal vortices. As explained by Wei and Smith \cite{Wei1986}, the secondary vortices are only due to a free shear instability in the boundary layer whereas Strouhal vortices also require a fluctuating base pressure near the rear stagnation point. These two processes lead to different Strouhal numbers, for which the ratio depends on the Reynolds number \cite{Thompson2005}.

Figure \ref{fig:FigDiscWidth}b shows that the circulation $\Gamma_s$ of secondary vortices increases linearly with the quantity $\sqrt{eR}/V_d$, equivalent to a circulation. For the full vortex rings generated by a linear disc motion, Taylor \cite{Taylor1953} predicted that the infinitesimal circulation $\delta \Gamma$ produced in the circular plane annulus is:
\begin{equation}
	\delta\Gamma=\frac{4V_{d}r}{\pi R}\left(1-\frac{r^{2}}{R^{2}}\right)^{-\frac{1}{2}}\delta r
	\label{DeltaCirc}
\end{equation}
We assume that we can apply equation \eqref{DeltaCirc} one the disk edge with a thickness $e\ll r$. Thus the circulation of a secondary vortex is:
\begin{equation}
	\Gamma_s=\int\limits _{R}^{R+e}\frac{4V_{d}r}{\pi R}\left(1-\frac{r^{2}}{R^{2}}\right)^{-\frac{1}{2}}dr=\frac{4RV_{d}}{\pi}\left(\left(1+\frac{e}{R}\right)^{2}-1\right)^{\frac{1}{2}}\approx\frac{4\sqrt{2}}{\pi}\sqrt{eR}V_{d}
	\label{CircSmallSVRs}
\end{equation}
This simple assumption predicts the linear increase of the circulation $\Gamma_s$ with the quantity $\sqrt{eR}/V_d$. Nevertheless the slope predicted by the model $4\sqrt{2}/\pi$ is five times greater than the slope experimentally measured ($\sim 0.36$). We assumed that the vorticity produced on the disk edge is similar to vorticity produced on the front face, that may be an oversimplified approximation. A better description of the boundary layer on the leading edge is required to predict the correct value of the dimensionless circulation $\Gamma_s/\sqrt{eR}V_d$ of the secondary vortices. 

\subsubsection{Dependence of the SVR final circulation}

We focus now on the final circulation $\Gamma_0$ of the main SVR reached after the formation process and the particular role of the disk thickness $e$.
As we mention above, Didden \cite{Didden1979} obtained the initial circulation of vortex rings generated by a piston stroke by calculating the vorticity flux through the piston orifice. For the SVR produced in our experiments, the vorticity flux is modified by the vortex shedding of the boundary layer formed on the front side of the disk.

We measure the SVR circulation $\Gamma_0$ for the three different thickness described above and a constant radius $R=10$ cm, after they have been released from the disk (see Figure \ref{fig:FigDiscWidth}).
\begin{figure}\centering
	\includegraphics[width=13cm,trim=0 0 0 0,clip=true]{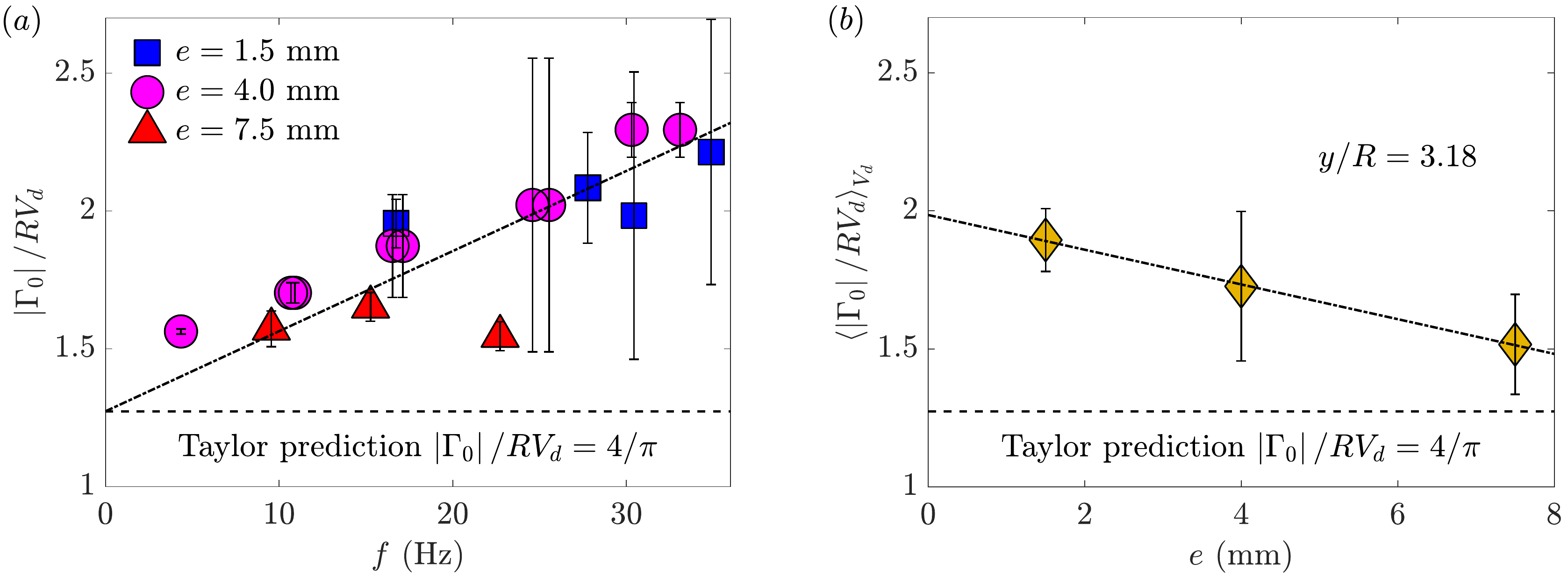}
	\caption{(a) Dimensionless circulations measured at a distance $y/R=3.18$ from the pendulum axis versus emission frequency $f$. The black dashed line indicates the dimensionless value $\Gamma_0/RV_d=4/\pi$ predicted by Taylor \cite{Taylor1953}. (b) Mean values of the dimensionless circulations for the three disk thickness Parameters: $L/R=3$, $R=10$ cm, $RV_d^{\mathrm{max}}/\nu=1.8\times 10^4$ and $v_m^{}=V_d^{}/V_d^{\mathrm{max}}$.}
	\label{fig:FigDiscWidth}
\end{figure}
In Figure \ref{fig:FigDiscWidth}a, we observe that the dimensionless circulation $\Gamma_0/RV_d$ increases when the emission frequency $f$ increases. It shows that the secondary vortices induced by the shedding process can merge like linear vortices \cite{Meunier2005}, and feed the main SVR generated at the rear of the disk. It is noteworthy that the dimensionless circulation value tends to the value $\Gamma_0/RV_d=4/\pi$ predicted by Taylor \cite{Taylor1953} for the linear displacement of a disk when the emission frequency tends to zero. In Figure \ref{fig:FigDiscWidth}b, we show that the mean values of the dimensionless circulation $\Gamma_0/RV_d$ for the three different thickness. The circulation decreases with increasing thickness and tends to $\Gamma_0/RV_d=4/\pi$ for large thickness values. In other words, the production of small SVR might be the main cause of the gap between our measurements and Taylor's prediction for normalized circulation. We assume that the disk thickness is involved in the final circulation of the main SVR through the shedding process. Notably, the circulation final value $\Gamma_0$ increases with decreasing thickness despite we observe that the circulation $\Gamma_s$ of the secondary vortices increases with decreasing thickness. Thus, the final circulation $\Gamma_0$ can not be assumed as a sum of the circulation $\Gamma_s$ of secondary vortices. The main SVR is the result of a complex merging process between the main SVR and several secondary vortices. Vortex merging is a complex process even for two vortices \cite{Meunier2005} and a model to predict the final circulation of a SVR after multiple merging process remains an open question.

\section{Conclusion}
We studied the formation and the scaling properties of a semi-vortex ring connected to a free surface generated by the circular motion of a thin flat disk, already observed for water strider. After the SVR has been released by the disk, we observe that this object is a tube with a constant vorticity and a semi-circular shape. The ring radius $h$ and the core size $\xi$ of the SVR are determined by the disk radius $R$ and obey to the equation \eqref{TaylorEq} given by Taylor \cite{Taylor1953}. Although a SVR is connected to a free surface, its displacement velocity $U_t$ and its circulation $\Gamma_0$ obey the same equation \eqref{UtLamb} given by Lamb \cite{Lamb1932} for a full vortex ring. A SVR slows down during its propagation but this decay might be also due to a vorticity transfer to the free surface and not only to the viscous dissipation as a full vortex ring. 

In first approximation, the circulation $\Gamma_0$ can be rescaled by the product $RV_d$ for disks of constant thickness. However, this simple scaling hides a complex formation process more complex than the vortex ring generated by a piston stroke. When the disk starts to move, the boundary layer separates and rolls up into a big SVR at the rear of the disk, similarly to the process occurring at the orifice outlet for vortex rings induced by a piston strokes. In the disk case, the boundary layer formed on its front side also separates but is destabilized by Kelvin-Helmholtz instability and periodically breaks on the disk edge, leading to a shedding phenomenon with emission of smaller satellite SVRs. The vortex shedding, taking place during the formation process, is due to a bi-dimensional instability called `transition waves' identified by Bloor \cite{Bloor1964} and can be enclosed in a Strouhal number $St=f\theta/V_d\approx 0.016$ where the characteristic length is the momentum thickness $\theta$ of the boundary layer formed on the disk edge. The secondary vortices have a dimensionless circulation $\Gamma_s$ proportional to $\sqrt{eR}V_d$ which can be explained by using Taylor \cite{Taylor1953} model.

Finally, we show that the secondary vortices are driven at the rear of the disk to finally merge and contribute to feed the main SVR. 
Thus the final circulation $\Gamma_0$ of the main SVR increases with the emission rate of secondary vortices. The disk thickness $e$ affects the final circulation $\Gamma_0$ through the emission frequency of secondary vortices produced. These information are essential to obtain a correct value of the initial circulation by calculating the vorticity flux and a better understanding of the locomotion of animals such as birds and fishes.

\begin{acknowledgments}

The authors would thank Amaury Fourgeaud and Olivier Brouard from the mechanic shop of PMMH for their technical support. The authors also thank Hamid Kellay, Joshua D. McGraw and Marion Matheli\'e-Guinlet for patient readings and their relevant remarks. A. Vilquin was supported by the Grant DYNAMONDE ANR-12-B509-0027-01 of the Agence Nationale de la Recherche.
\end{acknowledgments}

\bibliography{SVRbiblioVersion1}

\end{document}